\documentclass[12pt]{article}

\usepackage[pdftex]{graphicx}
\usepackage{times,fancyhdr,subfig}

\newfam\bboardfam
\font\tenbboard=msbm10  
 \font\sevenbboard=msbm7
   \font\fivebboard=msbm5 
\textfont\bboardfam=\tenbboard
\scriptfont\bboardfam=\sevenbboard
\scriptscriptfont\bboardfam=\fivebboard
\def\bboard{\fam\bboardfam\tenbboard}
\def\R{{\bboard R}}     

\newcommand{\ra}{\rightarrow}

\newcommand{\rr}[2]{\raisebox{-1.8ex}{$\;
{\stackrel{#1}{\stackrel{\textstyle \rightleftharpoons}{\scriptstyle #2}}}\;$}}

\newcommand{\sgr}[3]{#1 \stackrel{#2}{\ra} #3}

\newcommand{\lap}{\mbox{$\cal L$}}
\newcommand{\og}{\overline{G}}
\newcommand{\Rp}{\mbox{$\R_{> 0}$}}

\newcommand{\citep}{\cite}

\title{A linear elimination framework}

\author{Jeremy Gunawardena \\[0.5em]
\normalsize{Department of Systems Biology, Harvard Medical School}\\
\normalsize{200 Longwood Avenue, Boston, MA 02115, USA.}\\[0.2em]
\normalsize{{\tt jeremy@hms.harvard.edu}} \\
}

\date{}

\begin{document}

\maketitle

\begin{abstract}
Key insights in molecular biology, such as enzyme kinetics \cite{mm13}, protein allostery \cite{mwc,knf} and gene regulation \cite{ajs82}, emerged from quantitative analysis based on time-scale separation, allowing internal complexity to be eliminated and resulting in the well-known formulas of Michaelis-Menten, Monod-Wyman-Changeux and Ackers-Johnson-Shea. In systems biology, steady-state analysis has yielded eliminations that reveal emergent properties of multi-component networks \cite{sf09,mg07,mg09}. Here we show that these analyses of nonlinear biochemical systems are consequences of the same linear framework, consisting of a labelled, directed graph on which a Laplacian dynamics is defined, whose steady states can be algorithmically calculated. Analyses previously considered distinct are revealed as identical, while new methods of analysis become feasible. 
\end{abstract}

\clearpage

\section*{The linear framework}

Biological systems may sometimes be in steady state, as when synthesis or growth is balanced by degradation or loss. Typically, this holds only for a limited time period. It may also sometimes be reasonable to assume an explicit separation of time scales, in which a sub-system is operating fast compared to the rest of the system. The fast components may then be treated as if they are at steady state relative to the slow components. In either context, steady-state analysis is required. The framework introduced here provides a systematic way to calculate steady states, with broad applicability to biochemical systems.

\begin{figure}
\centering 
\includegraphics[viewport=0 50 590 776,width=\textwidth,height=0.7\textheight,keepaspectratio]{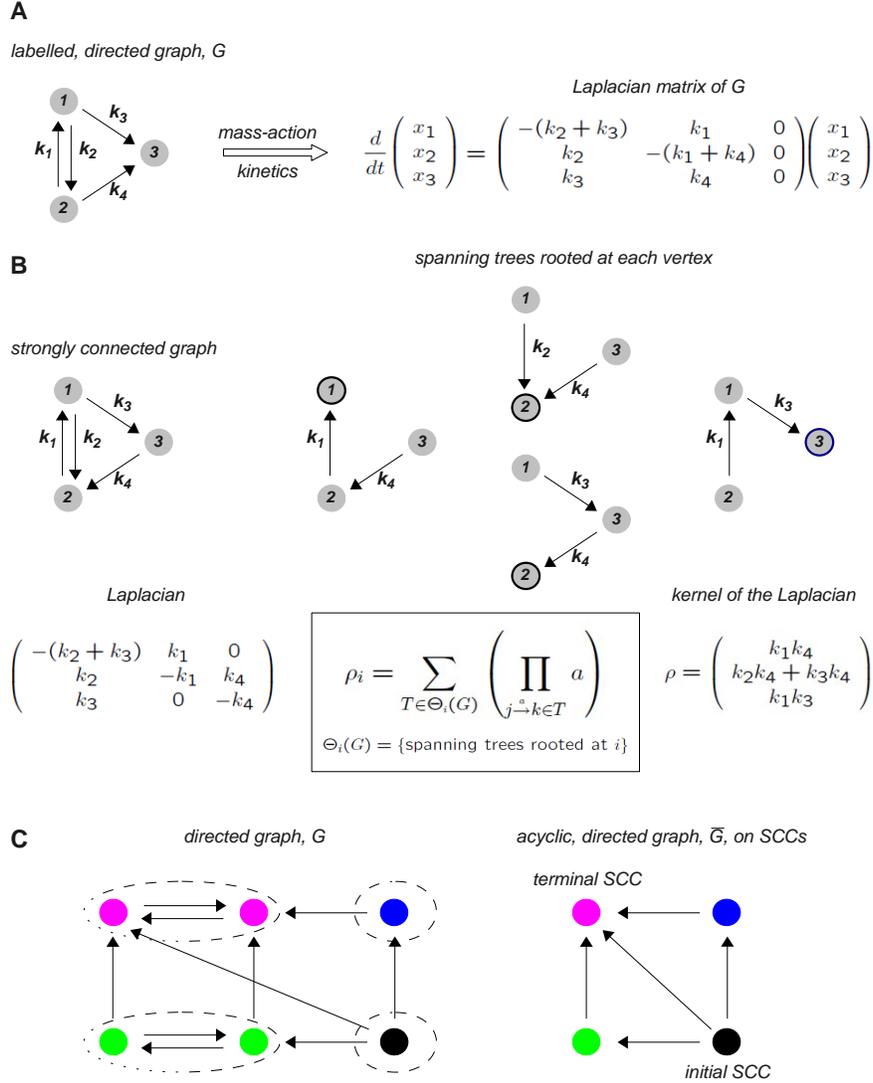}
\caption{The linear framework. {\bf (A)} A labelled, directed graph, $G$, gives rise to a system of linear differential equations by treating each edge as a first-order chemical reaction under mass-action kinetics, with the label as rate constant. The corresponding matrix is the Laplacian of $G$. {\bf (B)} In a strongly connected graph (note the difference to the one in {\bf A}), there are spanning trees rooted at each vertex, the roots being circled. The MTT gives an element of $\ker\lap(G)$ according to the formula in the box, as explained in the text. For more examples of spanning trees see Figures~1 and 2 of the Appendix. {\bf (C)} In a general directed graph, $G$, two distinct vertices are in the same strongly connected component (SCC) if each can be reached from the other by a path of directed edges. The SCCs form a directed graph, $\overline{G}$, in which two SCCs are linked by a directed edge if some vertex of the first SCC has an edge to some vertex of the second SCC. $\overline{G}$ has no directed cycles, allowing initial and terminal SCCs to be identified. \label{f-1}}
\end{figure}

We start from a graph, $G$, consisting of vertices, $1, \cdots, n$, with labelled, directed edges $\sgr{i}{a}{j}$ and no self loops, $i \not\ra i$ (Figure~\ref{f-1}A). The vertices represent components of a system, on which a dynamics is defined by treating each edge as if it were a first-order chemical reaction under mass-action kinetics, with the label as rate constant. This gives a system of linear, ordinary differential equations (ODEs),
\begin{equation}
\frac{dx}{dt} = \lap(G).x \,,
\label{e-lap}
\end{equation}
where $x$ is a column vector of component concentrations and $\lap(G)$ is the Laplacian matrix of $G$. Such matrices were introduced by Kirchhoff \cite{kir47} and resemble discretisations of the Laplacian operator (see the Appendix). 

Since material is neither created nor lost, the total concentration, $x_{tot} = x_1 + \cdots + x_n$, remains constant at all times, so that $1^{\dagger}.\lap(G) = 0$, where $1$ is the all-ones column vector and $^{\dagger}$ denotes transpose.

Nonlinearity can be encoded either in the vertices or, more commonly, in the labels. Labels are real numbers, $a \in \R$, which may be algebraic expressions over a set of symbols, $\{\mu_1, \cdots, \mu_s\}$. Symbols may be rate constants, $k$, or concentrations, $[X]$, of chemical species $X$.  For instance, $X$ may be a slow component in a time-scale separation. All calculations are in terms of symbols, whose numerical values do not have to be known in advance, thereby avoiding problems of parameter estimation. Labels must have dimensions of $\mbox{(time)}^{-1}$ and be positive, $a \in \Rp$. 
 
A crucial restriction is that if a concentration symbol, $[X]$, appears in a label in $G$, then $X$ must be an external species and not correspond to a vertex in $G$. This ``uncoupling condition'' is essential to preserve linearity and is the key requirement for applications of the framework. 

The Laplacian, $\lap(G)$, is a $n \times n$ matrix over $\R$. The interest lies in the steady states of (\ref{e-lap}), for which $dx/dt = 0$, or, equivalently, $x$ is in the kernel of the Laplacian, $x \in \ker\lap(G)$. The kernel can be determined in two steps, first for a strongly connected graph and then for any graph. 

A strongly connected graph is one in which any two distinct vertices can be joined by a series of edges in the same direction. While this depends only on the edge structure and not on the labels, the sign of a label determines the direction of flux. For strongly-connected graphs with positive labels, the dimension of $\ker\lap(G)$ is one,  \cite{mg09}. In this case, Tutte's Matrix-Tree Theorem (MTT) describes a basis element, $\rho \in \ker\lap(G)$,  \cite{tut48}. To calculate $\rho_i$, take the product of all the labels on a spanning tree of $G$ rooted at vertex $i$ and add the products over all such trees (Figure~\ref{f-1}B, box). A spanning tree is a fundamental concept in graph theory; it is a subgraph of $G$ that contains each vertex of $G$ (spanning) which has no cycles when edge directions are ignored (tree); it is rooted at $i$ if $i$ is the only vertex with no outgoing edges in the tree. Spanning-tree calculations are shown in Figure~\ref{f-1}B and Figures~1 and 2 of the Appendix. 

The kernel could have been calculated using determinants. The significance of the MTT is that it expresses $\rho_i$ as a polynomial in the labels with positive coefficients (Figure~\ref{f-1}B). This resolves the alternating signs that arise with determinants and ensures that steady-state concentrations remain positive, so long as the labels are positive. Being able to algorithmically calculate steady states in terms of labels is the essence of the framework. The MTT has been frequently rediscovered in biology in various guises,  \cite{ka56,hill66}.

If $x$ is any steady-state, then, since $\dim\ker\lap(G) = 1$, we know that $x = \lambda\rho$, where $\lambda \in \R$. The undetermined $\lambda$ reflects the amount of matter in the system. It can be removed by normalising in different ways:
\begin{equation}
1.\;\; x_i = \left(\frac{\rho_i}{\rho_1}\right)x_1 \hspace{4em} 2.\;\; x_i = \left(\frac{\rho_i}{\rho_{tot}}\right)x_{tot} \,.
\label{e-elim}
\end{equation}
In 1, one of the vertices, by convention vertex $1$, is chosen as a reference. In 2, $x_{tot}$ plays a similar role, with $\rho_{tot} = \rho_1 + \cdots + \rho_n$.

Equation (\ref{e-elim}) shows that the $n$ components in the system can be eliminated in favour of rational expressions, $\rho_i/\rho_1$ or $\rho_i/\rho_{tot}$, the labels of which may involve the concentrations of other components. This dramatic simplification is a consequence of strong connectivity and is central to the time-scale separation applications discussed below.

If $G$ is an arbitrary graph, it can be decomposed into strongly connected components (SCCs), which inherit from $G$ a directed graph structure, $\og$, that has no directed cycles (Figure~\ref{f-1}C). Since there is no net flux of material into the initial SCCs in $\og$, it can be shown that only the terminal SCCs contribute to any steady state (Appendix). For each terminal SCC, $t$, let $\rho^t \in \R^n$ be the vector which, for vertices in that SCC, agrees with the values coming from the MTT applied to that SCC in isolation, while for any other vertex, $j \not\in t$, $(\rho^t)_j = 0$. These vectors form a basis for the kernel of the Laplacian:
\begin{equation}
\ker\lap(G) = \langle\, \rho^1, \cdots, \rho^T \,\rangle \,,
\label{e-klap}
\end{equation}
where $T$ is the number of terminal SCCs. By construction, if $i$ is any vertex,
\begin{equation}
(\rho^t)_i \not= 0 \hspace{1em} \mbox{if, and only if, $i \in t$}
\label{e-rho}
\end{equation}
A description of $\ker\lap(G)$ appears in the Appendix of  \cite{fh77}. The construction given here goes further in using the MTT to give explicit expressions for the basis elements in terms of the labels.

Four applications are discussed next. The first stands apart from the rest in not being a time-scale separation. It illustrates the wide scope of the framework. The remaining applications show how the MTT systematises the eliminations arising from time-scale separation. In each case the framework integrates classical and modern analyses of biochemical systems. The intention is not to reveal new results in each area but to show that, rather than being different calculations, they are all the same calculation, made manifest in the labelled, directed graphs that appear in Figures~\ref{f-2} to \ref{f-5}. Following these applications, an extension to the framework is introduced that allows for synthesis and degradation of components (Figure~\ref{f-6}). Some specialised results for thermodynamic equilibrium are outlined in the Appendix.

\section*{Chemical Reaction Network Theory}

For a reversible chemical reaction between species $S_1, \cdots, S_k$ and species $P_1, \cdots, P_l$, 
\[ \alpha_1S_1 + \cdots + \alpha_kS_k \rr{k^+}{k^-} \beta_1P_1 + \cdots + \beta_lP_l \]
mass-action kinetics implies a Haldane relationship \cite{cb95} at equilibrium,
\begin{equation}
\frac{[S_1]^{\alpha_1} \cdots [S_k]^{\alpha_k}}{[P_1]^{\beta_1} \cdots [P_l]^{\beta_l}} = \frac{k^+}{k^-}\,.
\label{e-hald}
\end{equation}
Formula (\ref{e-hald}) may also be deduced from thermodynamics and, here, kinetics is consistent with thermodynamics. However, a network of reactions may have kinetic equilibria that do not satisfy thermodynamic constraints \cite{lew25}. The condition of ``detailed balance'' was introduced to avoid such paradoxes \cite{lew25,mah75}. This plays an important role at equilibrium, as explained in the next section.

\begin{figure}
\centering 
\includegraphics[viewport=30 56 778 568,width=\textwidth,height=\textheight,keepaspectratio]{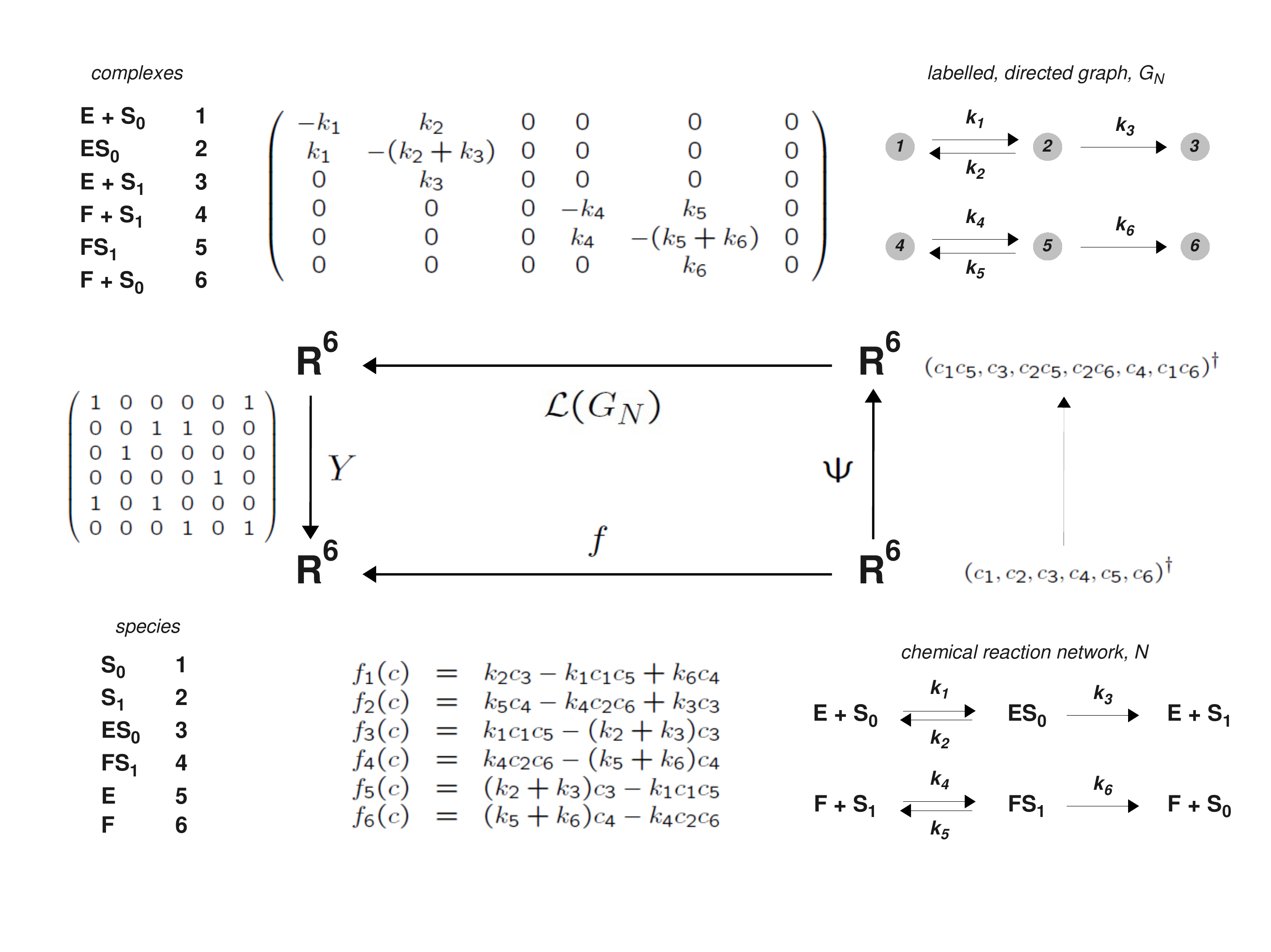}
\caption{Chemical Reaction Network Theory. A reaction network, $N$, is shown (bottom right) for a substrate $S$ existing in two states of modification, $S_0$ and $S_1$, which are inter-converted by enzymes $E$ and $F$. Each enzyme uses the classical Michaelis-Menten reaction mechanism, with enzyme-substrate complexes, $E\!S_0$ and $F\!S_1$. Mass-action kinetics gives a system of nonlinear differential equations, $dc/dt = f(c)$, where $c_i$ is the concentration of species $i$. The component functions $f_1(c), \cdots, f_6(c)$ are listed. The network gives rise to the labelled, directed graph $G_N$ on complexes (top right). The nonlinear function $f$ may be decomposed into the linear Laplacian, $\lap(G_N)$, as defined in Figure~\ref{f-1}A, and two linking functions, a linear function $Y$ and a nonlinear function $\Psi$. Formally, if $u_i$ denotes the multiplicity of species $i$ in complex $u$, then, for $c \in \R^m$, $\Psi(c)_u$ is the corresponding mass-action expression, $\Psi(c)_u = c_1^{u_1} \cdots c_n^{u_n}$ and if $z = (0, \cdots, 1, \cdots, 0)$ is the basis element of $\R^n$ corresponding to complex $u$, then $Y(z)$ is the list of multiplicities in $z$, $Y(z)_i = u_i$. With these definitions, the diagram in the centre commutes: $f(c) = Y\lap(G_N)\Psi$. \label{f-2}}
\end{figure}

In a seminal paper \cite{hj72}, Horn and Jackson, sought to extend thermodynamic properties like (\ref{e-hald}) to steady states far from equilibrium. Under mass-action kinetics, any reaction network gives rise to a system of nonlinear ODEs, $dc/dt = f(c)$. To disentangle the nonlinearity, the expressions that appear on either side of a reaction were treated as new entities called ``complexes'', so that a chemical reaction network, $N$, with $m$ species, gave rise to a labelled, directed graph, $G_N$, on $n$ complexes (Figure~\ref{f-2}). The nonlinear function $f$ on species is replaced by the linear Laplacian, $\lap(G_N)$, on complexes, with the labels being just the rate constants of the corresponding reactions. Here, the nonlinearity is entirely encoded in the vertices. (Horn and Jackson defined the function on complexes without being aware of its interpretation as a graph Laplacian.) The two functions, one acting on species and the other on complexes, are linked by a linear function $Y: \R^n \ra \R^m$ and a nonlinear function $\Psi: \R^m \ra \R^n$ (Figure~\ref{f-2}, caption). These encode the stoichiometry of the species in the complexes in such a way that that the diagram in Figure~\ref{f-2} commutes, $f(c) = Y\lap(G_N)\Psi(c)$. Only $\Psi$ is nonlinear, revealing a substantial linearity within the dynamics, arising from the graph-theoretic structure. This decomposition is the starting point of CRNT,  \cite{fein79,gun-cbp}.

Formula (\ref{e-klap}) applies to $\lap(G_N)$ and plays a fundamental role. If $c$ is positive, $c \in (\Rp)^n$, then, by definition, so is $\Psi(c) \in (\Rp)^m$. Hence, a positive steady state, with $f(c) = 0$, can arise in only one of two ways: either $\lap(G_N)\Psi(c) = 0$ or, if not, then $Y\lap(G_N)\Psi(c) = 0$. In the first case, $c$ is said to be ``complex balanced''. It then follows from (\ref{e-klap}) that
\[ \Psi(c) = \sum_{t=1}^T \lambda_t\rho^t \,,\]
where $\lambda_t \in \R$. Let $u$ and $v$ be two complexes in the same terminal SCC of $G_N$, say $t = t^*$. Suppose that the multiplicity of species $i$ in $u$ is $u_i$ and in $v$ is $v_i$. Using (\ref{e-rho}) and the definition of $\Psi$ (Figure~\ref{f-2}, caption), 
\begin{equation}
\frac{c_1^{u_1} \cdots c_n^{u_n}}{c_1^{v_1} \cdots c_n^{v_n}} = \frac{\Psi(c)_u}{\Psi(c)_v} = \frac{(\rho^{t^*})_u}{(\rho^{t^*})_v} \,.
\label{e-ghal}
\end{equation}
The term on the right depends only on the rate constants and this ``quasi-thermostatic'' property \cite{hj72} generalises the Haldane relationship in (\ref{e-hald}). With the MTT, the generalised ``equilibrium constants'' can now be explictly calculated in terms of the rate constants.

Horn and Jackson showed further that complex balancing satisifies other properties expected of thermodynamic equilibria, justifying it as a non-equilibrium generalisation of detailed balancing \cite{hj72}. 

Formula (\ref{e-klap}) has also provided modern insights. For instance, (\ref{e-ghal}) shows that complex-balanced steady states are generated by polynomials with only two terms (binomials),
\[ (\rho^{t^*})_v(c_1^{u_1} \cdots c_n^{u_n}) - (\rho^{t^*})_u(c_1^{v_1} \cdots c_n^{v_n}) = 0 \]
and therefore form a toric algebraic variety \cite{gh02,cdss08}, similar to those arising from log-linear models in algebraic statistics \cite{ascb}. This, and other recent results,  \cite{rg08}, have introduced methods of algebraic geometry to the analysis of molecular reaction networks.

Formula (\ref{e-klap}) remains useful even without a complex-balanced steady state. In the simplest case, $\ker Y\lap(G_N)$ contains only one additional basis element compared to $\ker\lap(G_N)$. If $c$ is a positive steady state, then
\[ \Psi(c) = \sum_{t=1}^T \lambda_t\rho^t + \lambda\chi \,,\]
where $\chi$ is the additional basis element and $\lambda \in \R$. Because $\chi$ may be non-zero at any complex, the Haldane-style formulas in (\ref{e-ghal}) can no longer be deduced, except in the case where $u$ and $v$ are not in any terminal SCC. If $u$ and $v$ also differ only in a single species $k$, so that $u_i = v_i$ for $i \not= k$, then $c_k$ depends only on the rate constants 
\[ c_k = \left(\frac{\chi_u}{\chi_v}\right)^{\frac{1}{u_k-v_k}} \]
and exhibits  ``absolute concentration robustness''. This is the Shinar-Feinberg Theorem \cite{sf09}, which has particular applications to bifunctional enzymes, where the robustness is suppored by experimental evidence  \cite{bg03,smma07,sra09}.

\section*{Reversible ligand binding}

\begin{figure}
\centering 
\includegraphics[viewport=38 0 744 568,width=0.9\textwidth,height=\textheight,keepaspectratio]{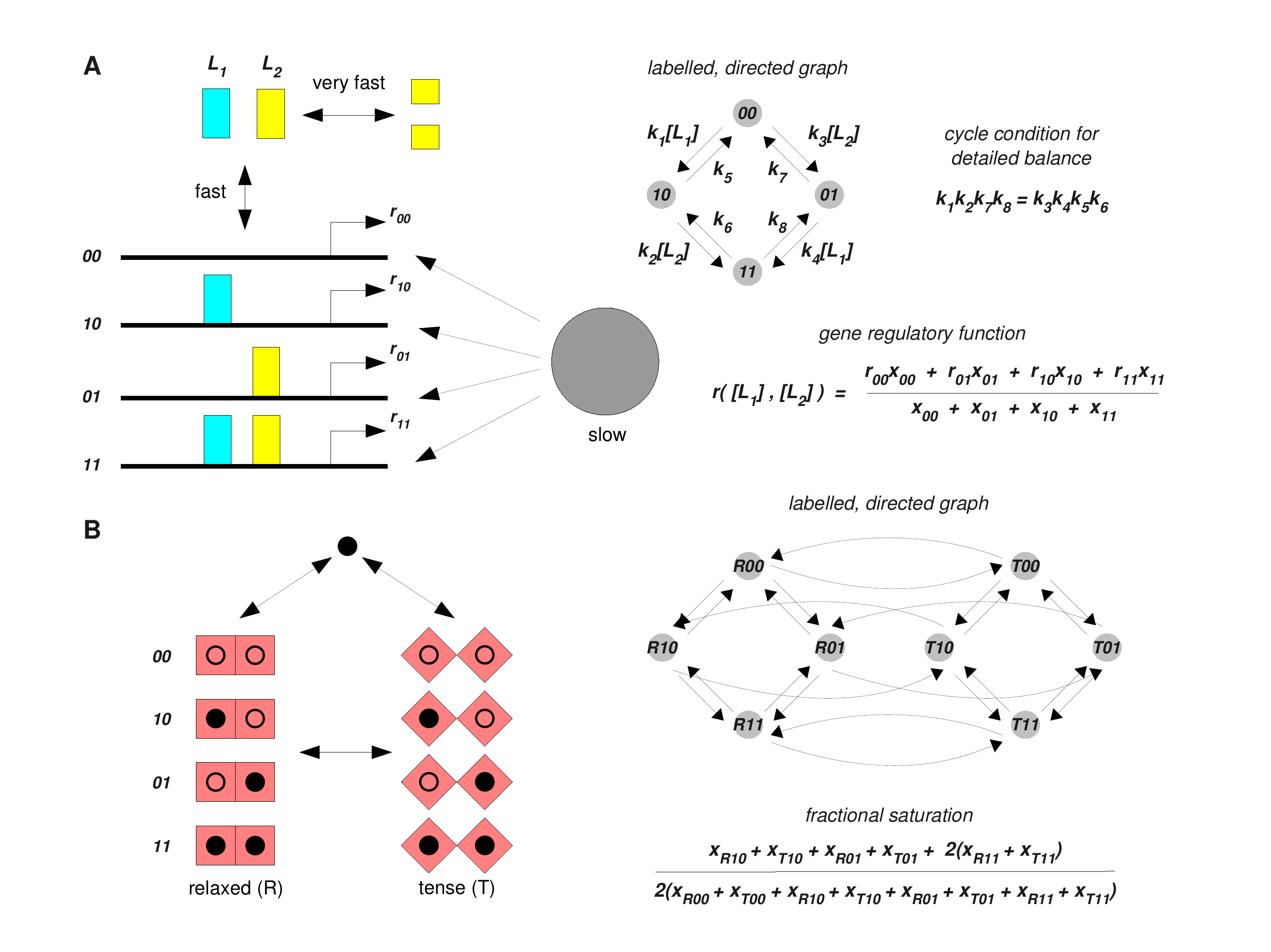}
\caption{Reversible ligand binding. {\bf (A)} Gene regulation,  \cite{ajs82,ssa03,bkr05-1,kor09}. Transcription factors may oligomerise before binding to DNA and initiating gene transcription at rates that depend on the pattern of ligand binding. A labelled, directed graph can be constructed as described in the text. Separating time scales as shown, the overall transcription rate as a function of oligomerised transcription factor concentrations (the gene regulation function), is the average rate, weighted by the probability of the promoter having the corresponding pattern of ligand binding. Probabilities are ratios $x_{00}/x_{tot}, \cdots, x_{11}/x_{tot}$ in any steady state $x$ of the Laplacian dynamics on the graph. In this example, reactions are asumed to take place at thermodynamic equilibrium, without dissipative changes, such as nucleosome reorganisation. {\bf (B)} Protein allostery,  \cite{mwc,nhm06}. An allosteric dimer is shown in two quaternary states, relaxed and tense. Ligand can bind to each monomer on a fast time-scale compared to catalytic activity of the protein. The labelled, directed graph has both quaternary state changes (relaxed to tense and vice versa) and ligand binding and unbinding, with the corresponding reactions being assumed to take place at themodynamic equilibrium. Labels have been omitted for clarity. For allosteric enzymes, the overall rate is assumed to be proportional to the fraction of sites that are bound by ligand (the fractional saturation), which can be directly calculated as described in the Appendix. \label{f-3}}
\end{figure}

Reversible binding of ligands to a substrate is a feature of many cellular processes, such as gene regulation,  \cite{ajs82}, and protein allostery,  \cite{mwc}. The linear framework may be readily applied by assuming that the time-scale of binding is well-separated between faster upstream interactions, such as ligand dimerisation, and slower downstream processes that react to the binding, such as gene expression (Figure~\ref{f-3}).

Consider a substrate that may exist in multiple states. These may, for instance, be states of DNA looping or nucleosome organisation at a promoter or conformational states in an allosteric protein. Ligands may bind reversibly to the substrate with potentially overlapping site preferences, cooperativity and dependence on substrate state. A labelled, directed graph can be constructed as follows (Figure~\ref{f-3}). The vertices correspond to microstates, consisting of the patterns of ligand binding in each substrate state. The edges correspond to transitions between substrate states, with ligand binding unaltered, or to binding or unbinding of the ligands, with substrate state unaltered. Of these edges, ligand binding has a label of the form $k[L]$, where $k$ is a rate constant and $[L]$ is a concentration, taken either at steady state or as slowly varying when $L$ is a slow variable; all other edges have only a rate constant as label. Provided the substrate is not a ligand for itself, so that the uncoupling condition is satisfied, and the graph is strongly connected, as is the case in most applications, the MTT allows the microstates to be eliminated in favour of the ligands. Most quantities of biological interest can be calculated in terms of the resulting expressions (Figure~\ref{f-3} and the Appendix).

An important special case is when the system can reach thermodynamic equilibrium (Appendix). In this case, detailed balance (DB) provides a simpler alternative to the MTT. According to DB, which follows from the fundamental reversibility of microscopic dynamics at equilibrium \cite{mah75}, each edge is reversible and any pair of reversible edges, $\sgr{i}{a}{j}$ and $\sgr{j}{b}{i}$, is independently at kinetic equilibrium. Hence, given any steady state $x$, $x_j = (a/b)x_i$, irrespective of any other edges that impinge on $i$ or $j$. Since each edge is reversible, the graph is strongly connected. Starting from a reference microstate, $1$, and taking a path of reversible edges to $j$, we find that $x_j = \alpha_j x_1$. Just as in (\ref{e-elim}), each $x_j$ can be eliminated in favour of rational expressions in the labels. At equilibrium, DB cuts down the rooted trees of the MTT to a single path from $1$.

There may be many such paths. However, the rate constants are not free to vary arbitrarily. DB requires that they yield the same $\alpha_j$ no matter what path is taken from $1$ to $j$. These constraints may be summarised in the ``cycle condition'': for any cycle of reversible edges, the product of the rate constants on clockwise edges equals the product on counterclockwise edges (Figure~\ref{f-3}A). This condition is necessary and sufficient for $\alpha_j$ to be independent of the path taken and for every equilibrium state to satisfy DB (Appendix).

Equilibrium ligand binding has usually been analysed by statistical mechanical methods,  \cite{hill85,wyman}, as in protein allostery,  \cite{mwc,nhm06}, and gene regulation,  \cite{ajs82,ssa03,bkr05-1}. The linear framework gives identical results from a more kinetic perspective. Its main advantge is that it also applies away from equilibrium. For instance, in the yeast phosphate control system, nucleosome organisation at the {\em PHO5} promoter influences its gene regulation function (GRF) in response to the transcription factor Pho4,  \cite{kos08}. Nucleation and disassembly of nucleosomes is a dissipative process. However, the GRF may still be calculated from the appropriate graph---Figure~4B in  \cite{kos08}---using the MTT. The linear framework is well suited to the modern programme of unravelling complex GRFs,  \cite{bkr05-1,kor09}

\section*{Enzyme kinetics}

The fundamental basis of enzymology is that enzymes act through intermediate enzyme-substrate complexes,  \cite{mm13,bc43}, (Figure~\ref{f-4}). Under {\em in-vitro} conditions, in which substrate is in excess, a time-scale separation may be assumed, with the intermediate complexes quickly reaching steady state, while conversion of substrate to product takes place more slowly. This is the quasi-steady state approximation, a version of which goes back to Michaelis and Menten,  \cite{mm13,cb95}. A labelled, directed graph can be constructed in which the vertices correspond to the intermediates and the free enzyme, with edges derived from the reaction mechanism. The labels can be chosen so that the differential equations of the linear Laplacian dynamics coincide with the full nonlinear ODEs. Since free substrate and free product are distinct from the intermediate complexes and the enzyme, the uncoupling condition is readily satisfied. Because intermediates eventually break up to release enzyme, the graphs are naturally strongly connected. The MTT and formula (\ref{e-elim}) can then be used to eliminate the intermediates and the free enzyme in favour of substrates and products, from which the enzymatic rate function can be calculated (Figure~\ref{f-4} and Appendix).

In the biochemical literature, such calculations are done by the King-Altman procedure,  \cite{ka56,cb95}, which is a restatement of the MTT. King-Altman has been widely used to calculate rate functions for complex enzymatic mechanisms with multiple ligands, affectors and intermediates,  \cite{seg93,cb95}. The linear framework both encompasses this and shows how it can be integrated into the analysis of multi-enzyme systems, as described next.

\section*{Post-translational modification (PTM)}

Many proteins are covalently modified by the attachment of small chemical or peptide moieties, such as phosphate or ubiquitin, to specific residues,  \cite{walsh}. PTM may involve multiple types of modifiers on multiple sites. Different global patterns of modification, or ``modforms'', may have different downstream effects, while the distribution of modforms is dynamically regulated by forward modifying and reverse demodifying enzymes acting in opposition,  \cite{psg11}, (Figure~\ref{f-5}). PTM is believed to implement adaptive cellular information processing on physiological time scales, as, for instance, in ``PTM codes'',  \cite{ja01,tur02}. The linear framework enables quantitative analysis despite the resulting dynamical and combinatorial complexity  \cite{mg07,mg09}.

Consider a single substrate, $S$, that supports multiple types of modification at multiple sites by multiple forward and reverse enzymes. Combinatorial explosion may lead to enormous numbers of modforms, depending on the numbers of sites and types of modification. A directed graph can be formed in which the vertices are the modforms and there is an edge between two modforms if there is some enzyme (there may be several) that catalyses the corresponding change in modification state. It is typically the case that any modification can be eventually undone by some other enzyme, so this modform graph is naturally strongly connected. 

The labels emerge from a separation of time scales. The donor molecules, such as ATP in the case of phosphorylation, and their breakdown products, such as ADP and phosphate, are assumed to be kept at constant concentration over the time scale of the modification dynamics by cellular processes that are not explicitly modelled. The modifier species can then be ignored as dynamical variables and enzyme reaction schemes can be simplified to involve only formation and breakdown of intermediate complexes and conversion between intermediate complexes (Figure~\ref{f-5}). Realistic enzyme mechanisms may be assumed that vary for different substrate modforms. The mechanisms can be analysed using the linear framework, as explained in the previous section, yielding expressions from which the labels for the modform graph can be assembled (Figure~\ref{f-5}, caption). The uncoupling condition becomes restrictive here, since it requires that no substrate is also a modifying or demodifying enzyme. The differential equations arising from the Laplacian dynamics then recapitulate the full nonlinear ODEs. 

Because the modform graph is strongly connected, the MTT can be applied to eliminate the modforms in favour of the enzymes. This is hierarchical elimination: the intermediates are first eliminated in favour of the modforms and the enzymes; the modforms are then eliminated in favour of the enzymes. We deduce that, despite the overwhelming combinatorial complexity arising from multisite modifications, the number of algebraically independent quantities at steady state is just the number of enzymes. This is usually very much smaller than the number of modforms. All other steady state concentrations are rational expressions in the free enzyme concentrations, with the expressions coming from (\ref{e-elim}). As for the enzymes, the total amount of each enzyme is conserved, which gives sufficiently many algebraic equations for the free enzyme values to be determined.

We see that the steady states of a PTM system can be calculated algebraically, without the need for numerical simulation, and without prior knowledge of any parameter values. This may be done irrespective of the number of modifications, the number of modification sites and the complex details of the enzyme mechanisms. 

This capability has yielded several insights,  \cite{mg07,rg08}. For instance, it has identified the first biochemical mechanism capable of implementing a ``PTM code'' and shown that its information capacity is potentially unlimited,  \cite{mg07}.

\section*{Synthesis and degradation}

The linear framework also provides a foundation for new types of analysis. An aspect of the applications above is that synthesis and degradation were ignored. This is tantamount to another assumption of time-scale separation, since cellular components are always being turned over. We now analyse what happens when this assumption is dropped.

Consider, as before, a labelled, directed graph, $G$, on vertices $1, \cdots, n$. Allow each vertex, $i$, to have a partial labelled edge leading in, $\sgr{\hspace{0.4em}}{a_i}{i}$, and out, $\sgr{i}{d_i}{\hspace{0.4em}}$, corresponding to zero-order synthesis or first-order degradation of $i$, respectively (Figure~\ref{f-6}A). By allowing $a_i = 0$ or $d_i = 0$, each vertex may have any combination of synthesis and degradation, including neither or both. The degradation label $d_i$ has the usual units of (time)$^{-1}$ but the synthesis label $a_i$ must have units of (concentration)(time)$^{-1}$. Call this ``partial graph'' $G^+$. As before, there is a linear dynamics on $G^+$, which may be described by the system of differential equations
\begin{equation}
\frac{dx}{dt} = \lap(G).x - \Delta.x + A \,.
\label{e-nlap}
\end{equation}
Here, $\Delta$ is a diagonal matrix with $\Delta_{ii} = d_i$ and $A$ is a column vector with $A_i = a_i$. Note that, unlike (\ref{e-lap}), the equations in (\ref{e-nlap}) are non-homogeneous: if $x$ is a steady state of (\ref{e-nlap}), it does not follow that $\lambda x$ is also a steady state. Because $1^\dagger.\lap(G) = 0$, if $x$ is a steady state of (\ref{e-nlap}), then
\begin{equation}
d_1x_1 + \cdots + d_nx_n = a_1 + \cdots + a_n
\label{e-sd}
\end{equation}
which reflects the fact that synthesis and degradation must be in overall balance.

When there is neither synthesis nor degradation, a general graph may have several degrees of freedom at steady state, reflected in the size of the basis in (\ref{e-klap}). These free quantities are ultimately determined by the initial conditions. With synthesis and degradation, some of these degree of freedom may be lost, as the total amount of matter is no longer conserved. This is reflected in the loss of homogeneity in (\ref{e-nlap}). The system may not reach a steady state unless synthesis and degradation can find a balance. 

Construct a new labelled, directed graph $G^*$ by adding a vertex $*$ to $G$ (Figure~\ref{f-6}B). For each partial edge $\sgr{\hspace{0.4em}}{a_i}{i}$ with $a_i > 0$ or $\sgr{i}{d_i}{\hspace{0.4em}}$ with $d_i > 0$, introduce the edges $\sgr{*}{a_i}{i}$ or $\sgr{i}{d_i}{*}$ in $G^*$, respectively. Unlike $G^+$, $G^*$ is a directed graph with positive labels, whose Laplacian dynamics are governed by (\ref{e-lap}).  It is easy to see that $(x_1, \cdots, x_n)$ is a steady state of $G^+$ if, and only, $(x_1, \cdots, x_n, 1)$ is a steady state of $G^*$. The condition for vertex $*$ to be at steady state in $G^*$ corresponds exactly to equation (\ref{e-sd}) for synthesis and degradation to be in balance in $G^+$.

This enables a complete description of the steady states of $G^+$ but we focus here on the case that is most relevant to the applications. If $G^*$ is strongly connected, so that the MTT gives $\rho$ as a basis element for the kernel of $\lap(G^*)$, then $G^+$ has a unique steady state $x$ for which
\begin{equation}
x_i = \frac{\rho_i}{\rho_*} \,.
\label{e-star}
\end{equation}
The single degree of freedom in $G^*$ has been used in (\ref{e-star}) to ensure that $x_* = 1$. Notice that $G^*$ may be strongly connected even though $G$ itself is not (Figure~\ref{f-6}), so that (\ref{e-star}) applies to a broader class of graphs than does the MTT itself.

Equation (\ref{e-star}) may be used to revisit the applications above to understand the impact of synthesis and degradation. It also opens up for analysis a broad range of new biological contexts. For instance, regulated degradation is a frequently used mechanism in several signal transduction pathways, such as the Wnt/beta-catenin and death-receptor pathways,  \cite{lskhk03,nle10}, which also make abundant use of reversible ligand binding and post-translational modification. Analysis of these using the linear framework is work in progress.

\section*{Conclusions}

Time-scale separation, leading to elimination of internal complexity, has been a fundamental method for analysing biochemical systems, from the earliest days of single-enzyme biochemistry through molecular biology to modern studies of multi-component systems. The framework shows that these calculations, which were previously considered distinct, are, in fact, the same. Moreover, they are all linear. The linearity hinges on the uncoupling condition, which allows nonlinearity in the dynamical variables to be traded for algebraic complexity in the labels. The fact that uncoupling is feasible in so many different contexts indicates a remarkable degree of linearity concealed within nonlinear biochemistry, a surprising insight that is amplified by the results of CRNT in Figure~\ref{f-2}. The framework brings systematic techniques, clarity and pedagogical coherence to the field and lays a foundation for developing new methods of analysis. 

One intriguing direction to explore is the extension of the framework from the steady state to the dynamics. The problem of whether time-scale separation yields a good approximation of the dynamics can be studied by the method of singular perturbation,  \cite{gun-cbp}. However, this has only been undertaken for a limited number of biological examples. The framework provides the means to formulate such an analysis in a far more general way.

In contrast to simulations, for which all details most be specified in advance, the framework yields results that hold irrespective of the underlying molecular complexity. It is, therefore, well suited for distilling biological principles without becoming mired in the molecular details, a much needed facility for modern biology.

\section*{APPENDIX}

\subsection*{Laplacian matrices and the MTT}

Matrices similar to the Laplacian in equation (\ref{e-lap}) were first introduced for unlabelled, undirected graphs by Gustav Kirchhoff in his 1847 paper,  \cite{kir47}, whose title, in English translation,  {\em ``On the solution of the equations obtained from the investigation of the linear distribution of galvanic currents''}, suggests its origins in his well-known studies of electrical circuits. In this form, the Laplacian may be seen as a discrete version of the continuous Laplacian operator but the same name is used for different versions and normalisations,  \cite{chung}. The concept of a spanning tree and a result similar to the Matrix Tree Theorem also make their appearance in Kirchhoff's paper. This seems to be the first of many subsequent Matrix Tree Theorems; see  \cite[Chapter~5]{moon} for historical references. Several deep properties of graphs emerge from the spectral theory of Laplacian matrices  \cite{chung}. Bill Tutte, one of the founders of modern graph theory, extended the concepts to directed graphs and proved the version of the MTT used here,  \cite{tut48}. 

\subsection*{Kernel of the Laplacian for a general graph}

We sketch a proof of equation (\ref{e-klap}) which gives a basis for the kernel of the Laplacian. While the essential ideas are introduced we leave it to the reader to fill in some of the details. Let $G$ be an arbitrary labelled, directed graph on the vertices, $1, \cdots, n$. As always, we assume that $G$ has no self loops. Choose $x \in \ker\lap(G)$. Let $\overline{G}$ be the acyclic directed graph on the strongly connected components (SCCs) of $G$, as in Figure~\ref{fs-1}C. Suppose that the vertices of $\overline{G}$ are $c_1, \cdots, c_m$ and that $c_1$ is an initial SCC that is not also terminal. By construction, there must be some vertex, $i_1 \in c_1$, with an edge leaving $c_1$, $i_1 \ra k$, where $k \not\in c_1$. If $x_{i_1} > 0$, there is a positive flux of material along this edge. For $x$ to be a steady state, this flux must be balanced by some flux coming into $i_1$. This can only arise from some edge $i_2 \ra i_1$ with $x_{i_2} > 0$. Taking all such vertices, recursively, yields a subset of vertices that can be the only source of the balancing flux into $i_1$. However, because $i_1$ is an initial SCC, this subset is entirely contained in $c_1$. Since this SCC has only a limited amount of material, it cannot indefinitely balance the outgoing flux on the edge $i_1 \ra k$. It follows that $x_{i_1} \leq 0$. However, if $x_{i_1} < 0$ then there is positive flux coming into $i_1$ along the edge $i_1 \ra k$. This can only be balanced by an edge $i_3 \ra i_1$ with $x_{i_3} < 0$. Arguing recursively in a similar way as above yields a similar contradiction. We conclude that $x_{i_1} = 0$. But then $x_j = 0$ for any vertex $j$ with $j \ra i_1$. Since $c_1$ is strongly connected, it is then easy to see that $x_j = 0$ for any $j \in c_1$. It follows that $x$ has no support on any initial SCC that is not also terminal. (The support of $x$ is the subset of vertices, $i$, such that $x_i \not= 0$.)

It is now easy to argue by induction over those SCCs that are not terminal to show that the support of $x$ contains only vertices that are in terminal SCCs. Consider each terminal SCC, $t$, as a labelled, directed graph, $G_t$, in its own right, in isolation from the rest of $G$. Assume that $G_t$ has $n_t$ vertices. Let $x^t \in \R^{n_t}$ be the vector obtained from $x$ by restricting $x$ to those vertices lying in $t$. Since $x$ has no support outside the terminal SCCs and there are no edges between the terminal SCCs, it should be clear that $x^t \in \ker\lap(G_t)$. Let $v^t \in \R^{n_t}$ is the vector coming from the MTT applied to $G^t$. Since $t$ is strongly connected and $\dim\ker\lap(G_t) = 1$, it must be that $x^t = \lambda^t v^t$, for some $\lambda^t \in \R$. Now let $\rho^t \in \R^n$ be the vector constructed for equation (\ref{e-klap}),
\[ (\rho^t)_i = \left\{
   \begin{array}{cl}
   (v^t)_i & \mbox{if $i \in t$} \\
   0 & \mbox{otherwise.}
   \end{array}\right.
\]
Since the terminal SCCs are disjoint, the vectors, $\rho^1, \cdots, \rho^T$, are linearly independent by construction. Evidently, $x = \sum_{t=1}^T \lambda^t\rho^t$. Hence, these vectors form a basis for the kernel of the Laplacian,
\[ \ker\lap(G) = \langle\, \rho^1, \cdots, \rho^T \,\rangle \,, \]
which proves equation (\ref{e-klap}).

\subsection*{Ligand binding at thermodynamic equilibrium}
\label{s-te}

Consider the labelled, directed graph, $G$, arising from the binding of multiple ligands to multiple sites on a substrate that may exist in multiple states, as discussed in the paper. The microstates are assumed to be encoded in some way, as in Figure~\ref{f-3}, and are enumerated simply as $1, \cdots, n$. Edges correspond either to changes in state of the substrate, with ligand binding unaltered, or to ligand binding or unbinding, with substrate state unaltered. Assuming that the system can reach thermodynamic equilibrium, each edge is reversible and edges can therefore be treated in pairs, 
\[ \sgr{i}{a^+_{i,j}}{j} \hspace{2em} \sgr{j}{a^-_{i,j}}{i} \]
A ligand binding edge is assumed to have a label, $a^+_{i,j} = k^+_{i,j}[L_u]$, where $k^+_{i,j}$ is a rate constant and $[L_u]$ is the concentration of one of the ligands, treated either at steady state or as slowly varying. For all other edges, the label is a rate constant.

If $x$ is a steady state of $G$---in other words, if $x \in \ker\lap(G)$---then $x$ satisfies DB if each reversible edge is independently at kinetic equilibrium. In other words, whenever there is a reversible edge, the forward and reverse fluxes are balanced,
\begin{equation}
a^+_{i,j}x_i = a^-_{i,j}x_j \,.
\label{e-kp}
\end{equation}
The cycle condition on $G$ states that, for any cycle of reversible edges, the product of the rate constants on the edges going clockwise is equal to the product of the rate constants on the edges going counterclockwise. We want to show that the cycle condition holds on $G$ if, and only, if every steady state satisfies DB.

Suppose first that $x$ satisfies DB. Since the net flux through any reversible edge is zero, the net flux around any cycle of reversible edges is also zero. We know from (\ref{e-kp}) that 
\begin{equation}
x_j = K_{i,j}x_i \,,
\label{e-kuv}
\end{equation}
where $K_{i,j} = a^+_{i,j}/a^-_{i,j}$. Choose any cycle of reversible edges and pick any two vertices on it, say $i'$ and $j'$. The cycle can be broken into a pair of directed paths from $i'$ to $j'$. Applying (\ref{e-kuv}) repeatedly on each path gives two expressions for $x_{j'}$ in terms of $x_{i'}$. Equating these expressions, cancelling ligand concentrations and clearing denominators, yields the cycle condition. Since the cycle was chosen arbitrarily, this proves the first part. 

Now suppose the cycle condition holds. Let $x$ be any steady state. We need to show that $x$ satisfies DB. We construct an alternative steady state $y$, which we show to satisfy DB, and then prove that $y = x$. Assume that the reference microstate, $1$, has no ligands bound, and set $y_1 = x_1$. For any other microstate $j$, choose some path of reversible edges from $1$ to $j$ and use (\ref{e-kuv}) to express $y_j$ in terms of $y_1$. Now choose some other path from $1$ to $j$ and obtain a second expression for $y_j$ in terms of $y_1$. The two paths together form a cycle of reversible edges, to which the cycle condition applies. Reorganising the cycle condition and putting in the appropriate ligand concentrations shows that the two path expressions give the same result for $y_j$. Hence, this quantity is well defined, irrespective of the path chosen.

We have unambiguously defined a state, $y$, of $G$ but we have yet to show that it is a steady state. Consider any reversible edge between the microstates $i$ and $j$. Choose a pair of reversible paths from $1$ to $i$ and from $1$ to $j$. Together with the reversible edge between $i$ and $j$, this gives a cycle of reversible edges. Applying the cycle condition, it is easy to see that, in the state $y$, the reversible edge between $i$ and $j$ must be in kinetic equilibrium. This not only implies that $y$ is a steady state but also that $y$ satisfies DB. But now, $G$ is strongly connected and so $\dim\ker\lap(G) = 1$. Hence, $y = \lambda x$ for some $\lambda \in \R$. Since $y_1 = x_1$, $\lambda = 1$. Hence, $y = x$ and therefore $x$ satisfies DB. This completes the proof.

If the reference vertex, $1$, has no ligands bound, then, in any steady state $x$, the quantity $x_i/x_1$ is a monomial in the ligand concentrations and the power to which $[L_u]$ appears is the number of $L_u$ molecules bound in microstate $i$. Hence, the concentration of states in which $L_u$ is bound is given by 
\[ [L_u](\partial x_{tot}/{\partial [L_u]}) \]
and the ``fractional saturation'', or average concentration of states bound by $L_u$, is the logarithmic derivative,
\begin{equation}
\left(\frac{[L_u]}{x_{tot}}\right)\frac{\partial x_{tot}}{\partial [L_u]} \,. 
\label{e-fs}
\end{equation}
More complex aggregate concentrations can be worked out in a similar way.

The calculation of $x_{tot}$ can be simplified by suitably decomposing the graph, as illustrated by the sum and product formulae below.

DB implies that any steady state $x^G$ of $G$ gives, by restriction, a steady state $x^R$ of any subgraph, $R$. If $R$ and $T$ are subgraphs that are disjoint (no vertex in common), which together span $G$, we get the sum formula
\begin{equation}
(x^G)_{tot} = (x^R)_{tot} + (x^T)_{tot} \,.
\label{e-sum}
\end{equation}

If ligands bind independently, so that the site-specific rate constants are independent of the microstate in which ligand binds, then the graph may be decomposed into a product of the graphs for single site binding. The product of two graphs is defined as follows. Suppose that $G$ is a labelled, directed graph on the vertices $g_1, \cdots, g_n$ and that $H$ is a labelled, directed graph on the vertices $h_1, \cdots, h_m$. The product $G \times H$ is the labelled, directed graph on the vertices $g_i \times h_j$ in which there is an edge
\[ \sgr{g_{i_1} \times h_{j_1}}{a}{g_{i_2} \times h_{j_1}} \]
whenever there is an edge $\sgr{g_{i_1}}{a}{g_{i_2}}$ in $G$ and, symmetrically, there is an edge
\[ \sgr{g_{i_1} \times h_{j_1}}{b}{g_{i_1} \times h_{j_2}} \]
whenever there is an edge $\sgr{h_{j_1}}{b}{h_{j_2}}$ in $H$. There are no edges in $G \times H$ other than these. This construction captures the fact that a change in state of either factor is independent of the state of the other factor. 

The steady state of a product may be obtained from those of its factors as follows. Define the normalised total steady state by $\pi(G) = x_{tot}/x_1$, where $x$ is any steady state. It follows from equation~\ref{e-elim} that $\pi(G)$ is independent of $x$, although it may depend on the choice of reference vertex. With $1 \times 1$ as the reference in $G \times H$, it is not difficult to prove the product formula,
\begin{equation}
\pi(G \times H) = \pi(G) \times \pi(H) \,.
\label{e-pi}
\end{equation}
Independent binding allows $\pi(G)$ to be factorised.

Formulae (\ref{e-fs}), (\ref{e-sum}) and (\ref{e-pi}) are helpful for the typical calculations arising in studies of gene regulation or protein allostery.

\subsection*{Enzyme kinetics}

\begin{figure}
\centering 
\includegraphics[viewport=58 94 526 536,width=\textwidth,height=0.4\textheight,keepaspectratio]{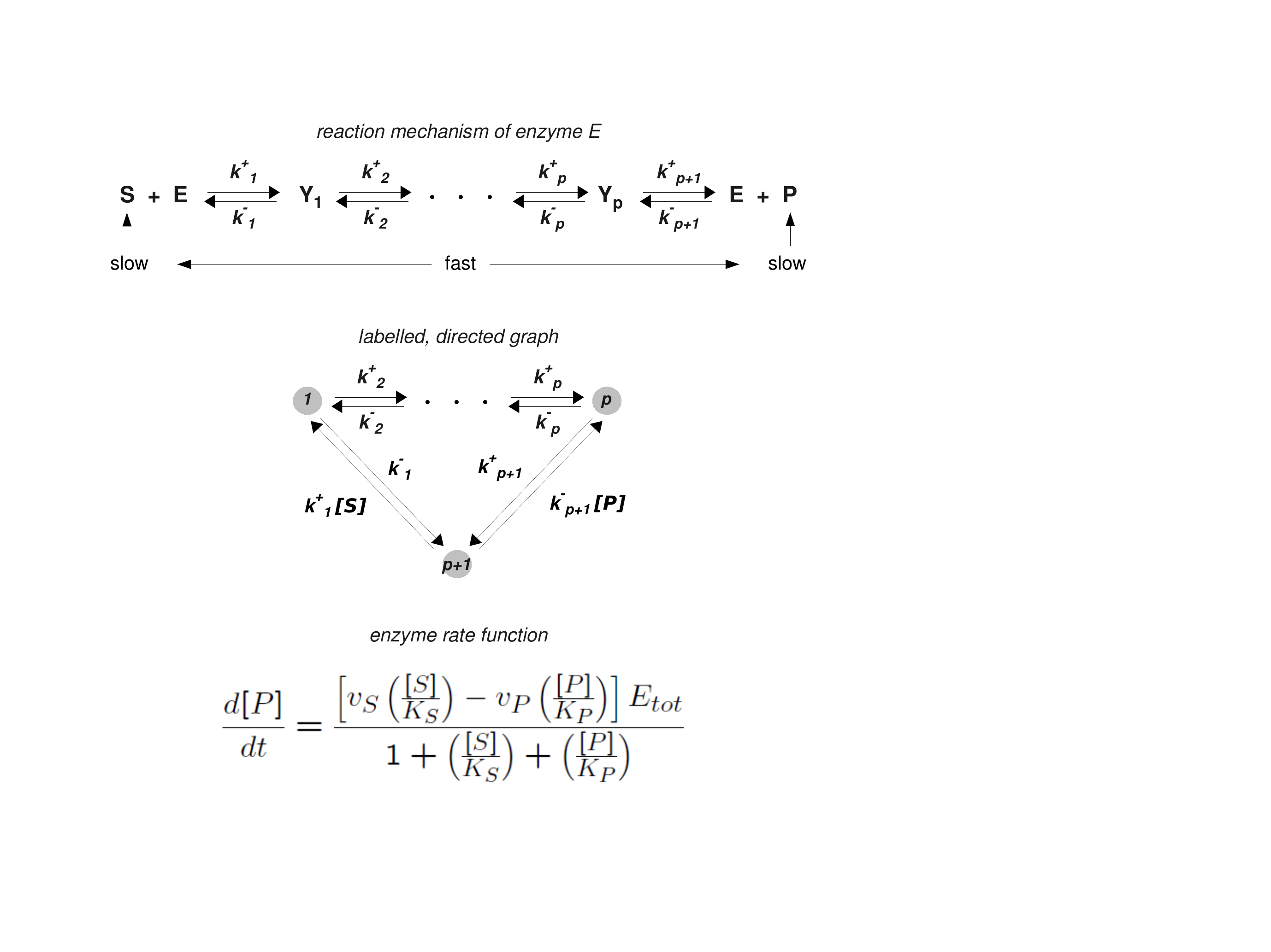}
\caption{Enzyme kinetics. A reaction mechanism is shown for enzyme $E$ converting substrate $S$ into product $P$ via the intermediate complexes $Y_1, \cdots, Y_p$. With the indicated separation of time scales, a labelled, directed graph can be contructed with vertices $1, \cdots, p$, corresponding to $Y_1, \cdots, Y_p$, respectively, and an additional vertex $p+1$, corresponding to the free enzyme, $E$. The two edges leading out of $p+1$, representing the formation of intermediate complexes, acquire algebraic expressions as labels, while all other edges have the corresponding rate constants. With this choice of labels the linear Laplacian dynamics on the graph recapitulates the full nonlinear dynamics of the reactions. The MTT can be used to be used to calculate the rate of product formation, $dP/dt = k^+_{p+1}[Y_p] - k^-_{p+1}[P][E]$, as explained in the SOM, leading to the reversible Michaelis-Menten formula,  \cite{cb95}.\label{f-4}}
\end{figure}

\begin{figure}
\centering 
\includegraphics[viewport=22 50 776 564,width=0.9\textwidth,height=\textheight,keepaspectratio]{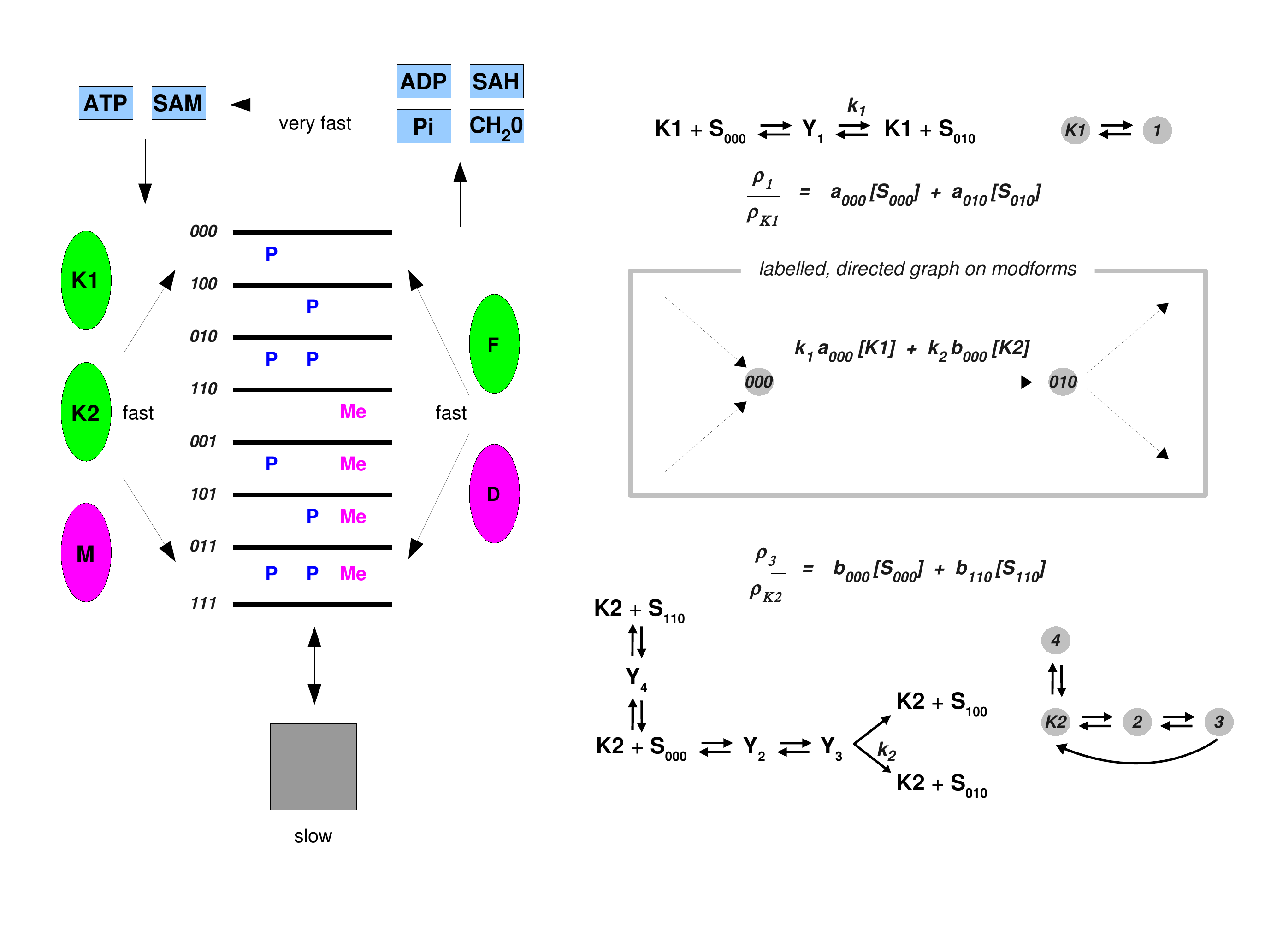}
\caption{Post-translational modification. A hypothetical example is shown with eight modforms of a substrate with two phosphorylation sites and one methylation site, acted on by kinases $K1$ and $K2$, methyltransferase $M$, phosphatase $F$ and demethylase $D$. The system is coupled upstream to core metabolism that renews the donor molecules (for methylation, SAM is S-adenosyl methionine, SAH is S-adenosyl homocysteine and CH$_2$O is formaldehyde) and downstream to the biological processes influenced by PTM . Assuming time-scale separations as shown, the PTM system gives rise to a directed graph on the modforms, one edge of which, from $000$ to $010$, is highlighted in the box. This edge is catalysed by $K1$ and $K2$ through the individual reaction mechanisms shown, in which the modifier species can be ignored because of the time-scale separation. $K1$ acts sequentially, producing only $010$ from $000$; $K2$ can produce both $010$ and $100$ in a random, distributive manner as well as the doubly phosphorylated $110$ processively. Also shown are the corresponding graphs on the intermediate complexes, constructed as in Figure~\ref{f-4}, with labels omitted for clarity. Using the MTT yields expressions whose coefficients, $a_{000}, a_{010}, b_{000}, b_{110}$, may be regarded as the reciprocals of generalised Michaelis-Menten constants. The appropriate label on the edge $\sgr{000}{}{010}$ can then be assembled as a linear combination of the steady-state concentrations of $K1$ and $K2$, with coefficients that are generalised catalytic efficiencies. See  \cite{mg09} for further details. \label{f-5}}
\end{figure}

\begin{figure}
\centering 
\includegraphics[viewport=18 42 778 558,width=0.9\textwidth,height=\textheight,keepaspectratio]{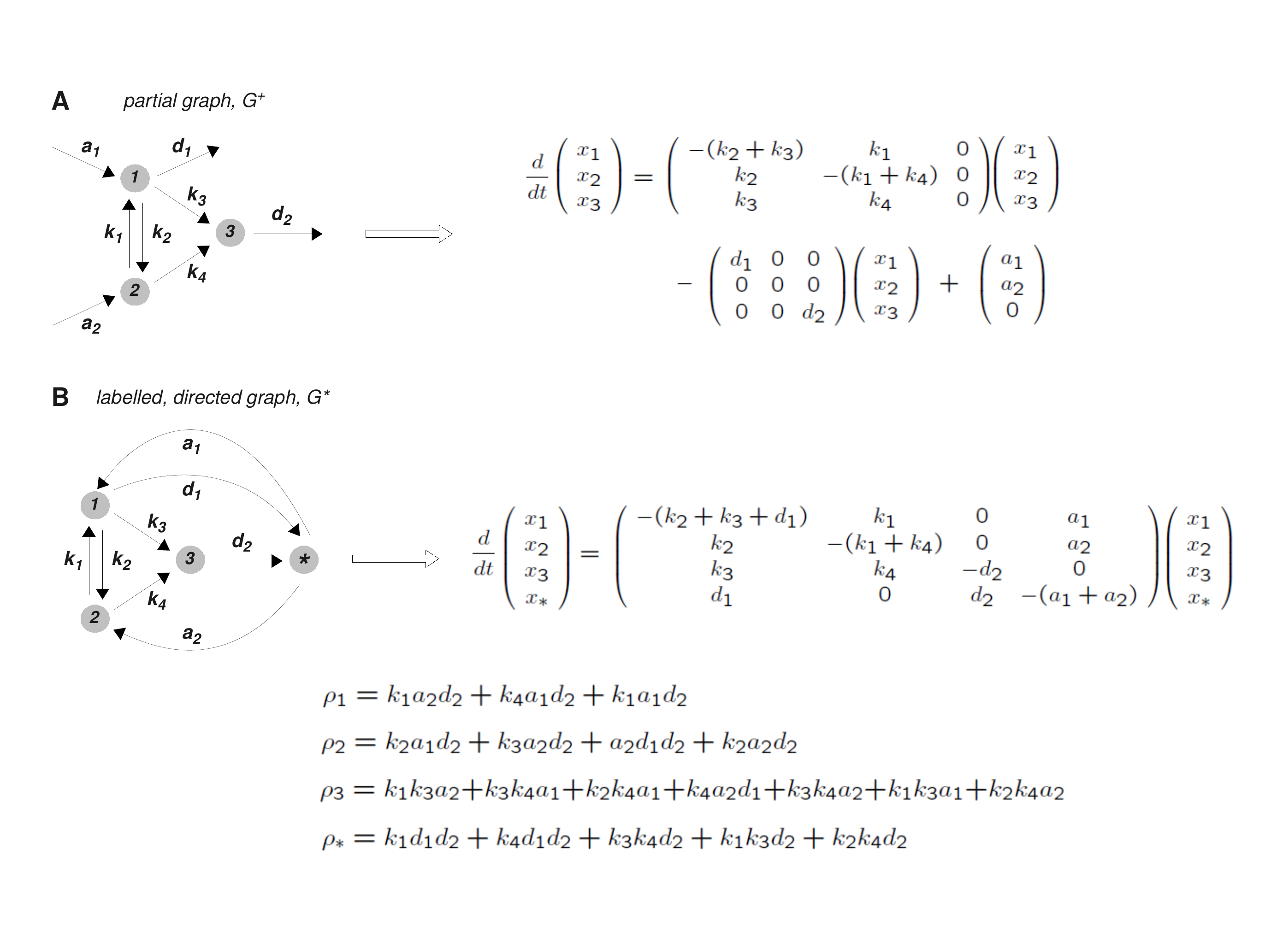}
\caption{Synthesis and degradation. {\bf (A)} The non-strongly connected graph in Figure~\ref{f-1}A is augmented with partial edges denoting synthesis and degradation to form the partial graph $G^+$. For clarity, only those partial edges with non-zero labels are shown. Under mass-action kinetics, $G^+$ gives rise to a non-homogenous system of linear ODEs. {\bf (B)} By introducing a new vertex, $*$, the labelled, directed graph, $G^*$,  can be formed, which, in this case, is strongly connected, with the corresponding Laplacian. Using the MTT to calculate $\rho \in \ker\lap(G^*)$, as shown (the spanning trees are enumerated in Figure~2 of the SOM), the unique steady state of $G^+$ can be calculated as $x_i = \rho_i/\rho_*$, for $1 \leq i \leq 3$. \label{f-6}}
\end{figure}

\begin{figure}
\centering 
\includegraphics[viewport=56 2 716 592,width=0.9\textwidth,height=\textheight,keepaspectratio]{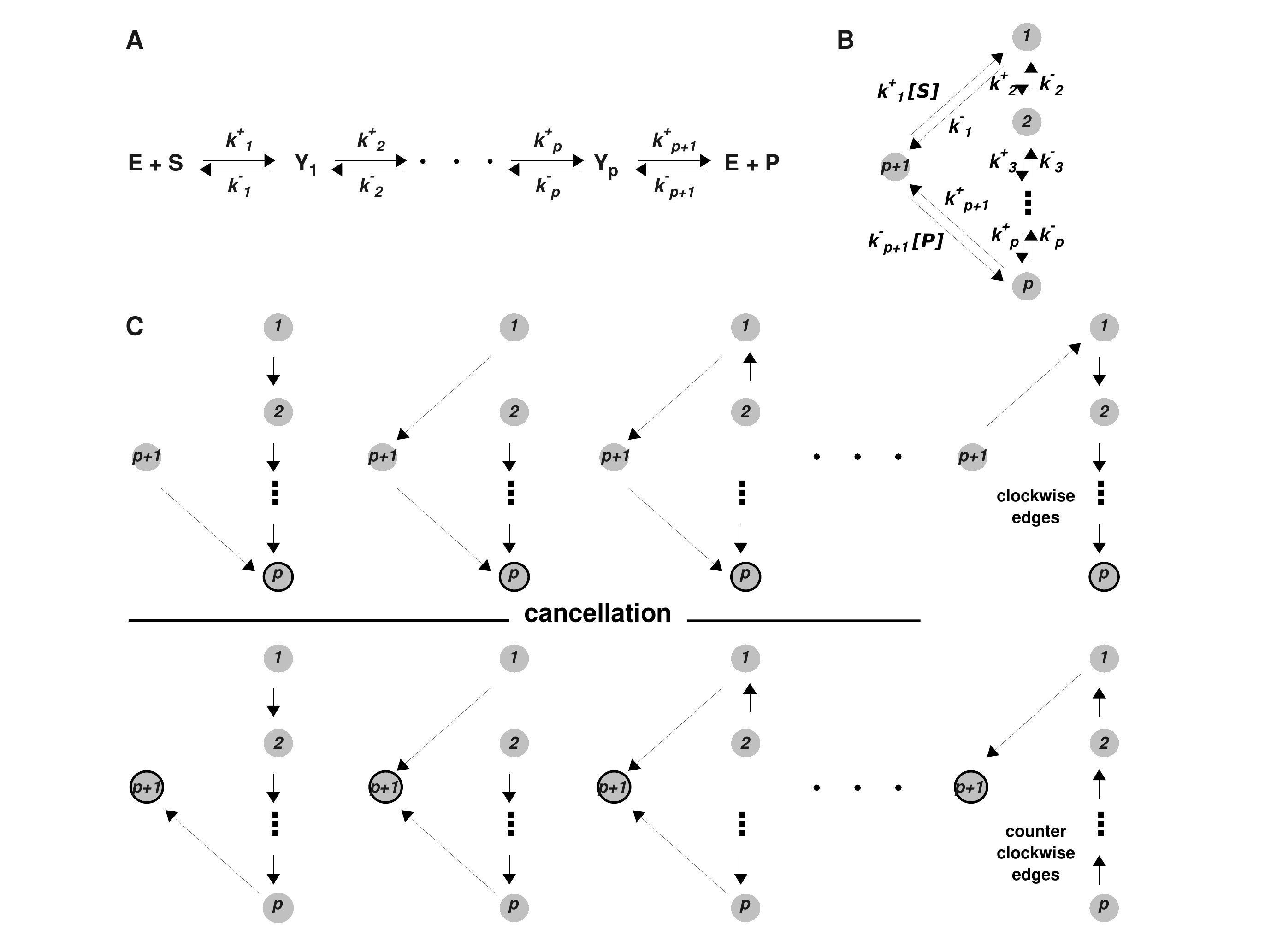}
\caption{Enzyme kinetics. {\bf (A)} The reaction mechanism from Figure~\ref{f-4}. {\bf (B)} The corresponding labelled, directed graph. Note that its orientation is different from that shown in Figure~\ref{f-4}. Note that only the two outgoing edges from vertex $p+1$ have algebraic expressions for labels. {\bf (C)}. Enumeration of the spanning trees rooted at vertex $p$ (top), corresponding to $Y_p$, and vertex $p+1$ (bottom), corresponding to $E$. The spanning trees for any root may be constructed by choosing a gap between adjacent vertices, with the edges running in opposite directions to the root on either side of the gap. Using the labels in {\bf B} and the MTT formula from Figure~\ref{f-1}B, it can be checked that only $\rho_{p+1}$ has no occurrences of $[S]$ or $[P]$, while each $\rho_i$, for $i \not= p+1$, has either one $[S]$ or one $[P]$ but not both. This proves equation (\ref{e-rt}). Considering the depicted trees in vertical pairs, it can similarly be checked that the pre-factor in equation (\ref{e-cir}) reduces to the two terms coming from the last pair of trees, giving the formula in equation (\ref{e-cw}). \label{fs-1}}
\end{figure}

The details of the calculation of the enzymatic rate formula in Figure~\ref{f-4} are shown in Figure~\ref{fs-1}. The rate of product formation is given by
\begin{equation}
\frac{d[P]}{dt} = k^+_{p+1}[Y_p] - k^-_{p+1}[P][E] \,.
\label{e-ert}
\end{equation}
Using the ordering in Figure~\ref{fs-1}, in which vertex $p+1$ corresponds to $E$, the elimination formula in equation (\ref{e-elim}) gives $[Y_p] = (\rho_p/\rho_{tot})E_{tot}$ and $[E] = (\rho_{p+1}/\rho_{tot})E_{tot}$. Hence,
\begin{equation}
\frac{d[P]}{dt} = (k^+_{p+1}\rho_p - k^-_{p+1}[P]\rho_{p+1})\left(\frac{E_{tot}}{\rho_{tot}}\right) \,.
\label{e-cir}
\end{equation}
The spanning trees of an isolated cycle are easily enumerated (Figure~\ref{fs-1}C) and the MTT shows that $\rho_{p+1} = \gamma$ and $\rho_i = \alpha_i[S]+\beta_i[P]$, for $i < p$, where $\gamma, \alpha_i, \beta_i$ are polynomials in the rate constants. Hence,
\begin{equation}
\rho_{tot} = \gamma + \left(\sum_{i=1}^p \alpha_i\right)[S] + \left(\sum_{i=1}^p \beta_i\right)[P] \,.
\label{e-rt}
\end{equation}
Comparing the spanning trees for vertices $p$ and $p+1$ reveals substantial cancellation when calculating the pre-factor in (\ref{e-cir}) (Figure~\ref{fs-1}C). This simplifies to the difference between the product of the labels going clockwise around the cycle and the product of the labels going counterclockwise,
\begin{equation}
k^+_{p+1}\rho_p - k^-_{p+1}[P]\rho_{p+1} = \underbrace{k^+_1[S]k^+_2 \cdots k^+_{p+1}}_{\mbox{\small labels on CW edges}} \;\; - \;\; \underbrace{k^-_1k^-_2 \cdots k^-_{p+1}[P]}_{\mbox{\small labels on CCW edges}} \,.
\label{e-cw}
\end{equation}
Note that the term on the right in (\ref{e-ert}) is the steady-state net flux around the isolated cycle in Figure~\ref{fs-1}B. When this is zero, (\ref{e-cw}) shows that the product of the clockwise labels equals the product of the counterclockwise labels. This gives another proof of the cycle condition, discussed in \S\ref{s-te}, which holds at thermodynamic equilibrium.

Combining (\ref{e-rt}) and (\ref{e-cw}) and normalising appropriately yields the rate formula
\begin{equation}
\frac{d[P]}{dt} = \frac{\left[v_S\left(\frac{[S]}{K_S}\right) - v_P\left(\frac{[P]}{K_P}\right)\right]E_{tot}}{1 + \left(\frac{[S]}{K_S}\right) + \left(\frac{[P]}{K_P}\right)} \,,
\label{e-mmr}
\end{equation}
where $v_S$, $v_P$, $K_S$, $K_P$ are rational expressions in the rate constants. This is the reversible Michaelis-Menten formula \cite{cb95}. Note that this has the same form irrespective of the number of intermediates.


\begin{thebibliography}{10}

\bibitem{ajs82}
G.~K. Ackers, A.~D. Johnson, and M.~A. Shea.
\newblock Quantitative model for gene regulation by lambda phage repressor.
\newblock {\em Proc. Natl. Acad. Sci. USA}, 79:1129--33, 1982.

\bibitem{bg03}
E.~Batchelor and M.~Goulian.
\newblock Robustness and the cycle of phosphorylation and dephosphorylation in
  a two-component regulatory system.
\newblock {\em Proc. Natl. Acad. Sci. USA}, 100:691--6, 2003.

\bibitem{bkr05-1}
L.~Bintu, N.~E. Buchler, G.~G. Garcia, U.~Gerland, T.~Hwa, J.~Kondev, and
  R.~Phillips.
\newblock Transcriptional regulation by the numbers: models.
\newblock {\em Curr. Opin. Gen. Dev.}, 15:116--24, 2005.

\bibitem{bc43}
B.~Chance.
\newblock The kinetics of the enzyme-substrate compound of peroxidase.
\newblock {\em J. Biol. Chem.}, 151:553--77, 1943.

\bibitem{chung}
F.~R.~K. Chung.
\newblock {\em Spectral Graph Theory}.
\newblock Number~92 in Regional Conference Series in Mathematics. American
  Mathematical Society, 1997.

\bibitem{cb95}
A.~Cornish-Bowden.
\newblock {\em Fundamentals of Enzyme Kinetics}.
\newblock Portland Press, London, UK, 2nd edition, 1995.

\bibitem{cdss08}
G.~Craciun, A.~Dickenstein, A.~Shiu, and B.~Sturmfels.
\newblock Toric dynamical systems.
\newblock {\em J. Symb. Comp.}, 44:1551--65, 2009.

\bibitem{fein79}
M.~Feinberg.
\newblock Lectures on {C}hemical {R}eaction {N}etworks.
\newblock Lecture notes, {M}athematics {R}esearch {Center}, {U}niversity of
  {W}isconsin, 1979.

\bibitem{fh77}
M.~Feinberg and F.~Horn.
\newblock Chemical mechanism structure and the coincidence of the
  stoichiometric and kinetic subspace.
\newblock {\em Arch. Rational Mech. Anal.}, 66:83--97, 1977.

\bibitem{gh02}
K.~Gatermann and B.~Huber.
\newblock A family of sparse polynomial systems arising in chemical reaction
  systems.
\newblock {\em J. Symbolic Computation}, 33:273--305, 2002.

\bibitem{gun-cbp}
J.~Gunawardena.
\newblock Modelling of interaction networks in the cell: theory and
  mathematical methods.
\newblock In E.~Egelmann, editor, {\em Comprehensive Biophysics}, volume~9.
  Elsevier, 2011.

\bibitem{hill66}
T.~L. Hill.
\newblock Studies in irreversible thermodynamics {IV}. {D}iagrammatic
  representation of steady state fluxes for unimolecular systems.
\newblock {\em J. Theoret. Biol.}, 10:442--59, 1966.

\bibitem{hill85}
T.~L. Hill.
\newblock {\em Cooperativity Theory in Biochemistry: Steady-State and
  Equilibrium Systems}.
\newblock Springer Series in Molecular Biology. Springer-Verlag, New York, USA,
  1985.

\bibitem{hj72}
F.~Horn and R.~Jackson.
\newblock General mass action kinetics.
\newblock {\em Arch. Rational Mech. Anal.}, 47:81--116, 1972.

\bibitem{ja01}
T.~Jenuwein and C.~D. Allis.
\newblock Translating the histone code.
\newblock {\em Science}, 293:1074--80, 2001.

\bibitem{kos08}
H.~D. Kim and E.~K. O'Shea.
\newblock A quantitative model of transcription factor-activated gene
  expression.
\newblock {\em Nat. Struct. Mol. Biol.}, 15:1192--8, 2008.

\bibitem{kor09}
H.~D. Kim, T.~Shay, E.~K. O'Shea, and A.~Regev.
\newblock Transcriptional regulatory circuits: predicting numbers from
  alphabets.
\newblock {\em Science}, 325:429--32, 2009.

\bibitem{ka56}
E.~L. King and C.~Altman.
\newblock A schematic method of deriving the rate laws for enzyme-catalyzed
  reactions.
\newblock {\em J. Phys. Chem.}, 60:1375--8, 1956.

\bibitem{kir47}
G.~Kirchhoff.
\newblock {\"U}ber die {A}ufl{\"o}sung der {G}leichungen, auf welche man bei
  der {U}ntersuchung der linearen {V}erteilung galvanischer {S}tr{\"o}me
  gef{\"u}hrt wird.
\newblock {\em Ann. Phys. Chem.}, 72:497--508, 1847.

\bibitem{knf}
D.~E. Koshland, G.~N{\'e}methy, and D.~Filmer.
\newblock Comparison of experimental binding data and theoretical models in
  proteins containing subunits.
\newblock {\em Biochemistry}, 5:365--85, 1966.

\bibitem{lskhk03}
E.~Lee, A.~Salic, R.~Kruger, R.~Heinrich, and M.~W. Kirschner.
\newblock The roles of {APC} and {A}xin derived from experimental and
  theoretical analysis of the {W}nt pathway.
\newblock {\em PLoS Biol.}, 1:116--32, 2003.

\bibitem{lew25}
G.~N. Lewis.
\newblock A new principle of equilibrium.
\newblock {\em Proc. Natl. Acad. Sci. USA}, 11:179--83, 1925.

\bibitem{mah75}
B.~H. Mahan.
\newblock Microscopic reversibility and detailed balance.
\newblock {\em J. Chem. Educ.}, 52:299--302, 1975.

\bibitem{rg08}
A.~Manrai and J.~Gunawardena.
\newblock The geometry of multisite phosphorylation.
\newblock {\em Biophys. J.}, 95:5533--43, 2008.

\bibitem{mm13}
L.~Michaelis and M.~Menten.
\newblock Die kinetik der {I}nvertinwirkung.
\newblock {\em Biochem. Z.}, 49:333--69, 1913.

\bibitem{mwc}
J.~Monod, J.~Wyman, and J.~P. Changeux.
\newblock On the nature of allosteric transitions: a plausible model.
\newblock {\em J. Mol. Biol.}, 12:88--118, 1965.

\bibitem{moon}
J.~W. Moon.
\newblock {\em Counting Labelled Trees}.
\newblock Number~1 in Canadian Mathematical Monographs. Canadian Mathematical
  Congress, 1970.

\bibitem{nhm06}
T.~S. Najdi, C.~R. Yang, B.~E. Shapiro, G.~W. Hatfield, and E.~D. Mjolsness.
\newblock Application of a generalised {MWC} model for the mathematical
  simulation of metabolic pathways regulated by allosteric enzymes.
\newblock {\em J. Bioinform. Comput. Biol.}, 4:335--55, 2006.

\bibitem{nle10}
L.~Neumann, C.~Pforr, J.~Beaudoin, A.~Pappa, N.~Fricker, P.~H. Krammer, I.~N.
  Lavrik, and R.~Eils.
\newblock Dynamics within the {CD95} death-inducing signaling complex decide
  life and death of cells.
\newblock {\em Mol. Syst. Biol.}, 6:352, 2010.

\bibitem{ascb}
L.~Pachter and B.~Sturmfels, editors.
\newblock {\em Algebraic Statistics for Computational Biology}.
\newblock Cambridge University Press, 2005.

\bibitem{psg11}
S.~Prabhakaran, R.~A. Everley, I.~Landrieu, J.~M. Wieruszeski, G.~Lippens,
  H.~Steen, and J.~Gunawardena.
\newblock Comparative analysis of erk phosphorylation suggests a mixed strategy
  for measuring phospho-form distributions.
\newblock {\em Mol. Sys. Biol.}, 7:482, 2011.

\bibitem{seg93}
I.~H. Segel.
\newblock {\em Enzyme Kinetics: Behaviour and Analysis of Rapid Equilibrium and
  Steady-State EnzymeSystems}.
\newblock Wiley-Interscience, 1993.

\bibitem{ssa03}
Y.~Setty, A.~E. Mayo, M.~G. Surette, and U.~Alon.
\newblock Detailed map of a cis-regulatory input function.
\newblock {\em Proc. Natl. Acad. Sci. USA}, 100:7702--7, 2003.

\bibitem{sf09}
G.~Shinar and M.~Feinberg.
\newblock Structural sources of robustness in biochemical networks.
\newblock {\em Science}, 327:1389--91, 2010.

\bibitem{smma07}
G.~Shinar, R.~Milo, M.~R. Mart{\'i}nez, and U.~Alon.
\newblock Input-output robustness in simple bacterial signaling systems.
\newblock {\em Proc. Natl. Acad. Sci. USA}, 104:19931--5, 2007.

\bibitem{sra09}
G.~Shinar, J.~D. Rabinowitz, and U.~Alon.
\newblock Robustness in glyoxylate bypass regulation.
\newblock {\em PLoS Comp. Biol.}, 5:e1000297, 2009.

\bibitem{mg09}
M.~Thomson and J.~Gunawardena.
\newblock The rational parameterisation theorem for multisite
  post-translational modification systems.
\newblock {\em J. Theor. Biol.}, 261:626--36, 2009.

\bibitem{mg07}
M.~Thomson and J.~Gunawardena.
\newblock Unlimited multistability in multisite phosphorylation systems.
\newblock {\em Nature}, 460:274--7, 2009.

\bibitem{tur02}
B.~Turner.
\newblock Cellular memory and the histone code.
\newblock {\em Cell}, 111:285--91, 2002.

\bibitem{tut48}
W.~T. Tutte.
\newblock The dissection of equilateral triangles into equilateral triangles.
\newblock {\em Proc. Camb. Phil. Soc.}, 44:463--82, 1948.

\bibitem{walsh}
C.~T. Walsh.
\newblock {\em Posttranslational Modification of Proteins}.
\newblock Roberts and Company, Englewood, Colorado, 2006.

\bibitem{wyman}
J.~Wyman and S.~J. Gill.
\newblock {\em Binding and Linkage: Functional Chemistry of Biological
  Macromolecules}.
\newblock University Science Books, 1990.

\end{thebibliography}
\bibliographystyle{plain}

\newpage

\begin{figure}
\centering 
\subfloat{\includegraphics[viewport=32 36 756 586,width=0.8\textwidth,height=\textheight,keepaspectratio]{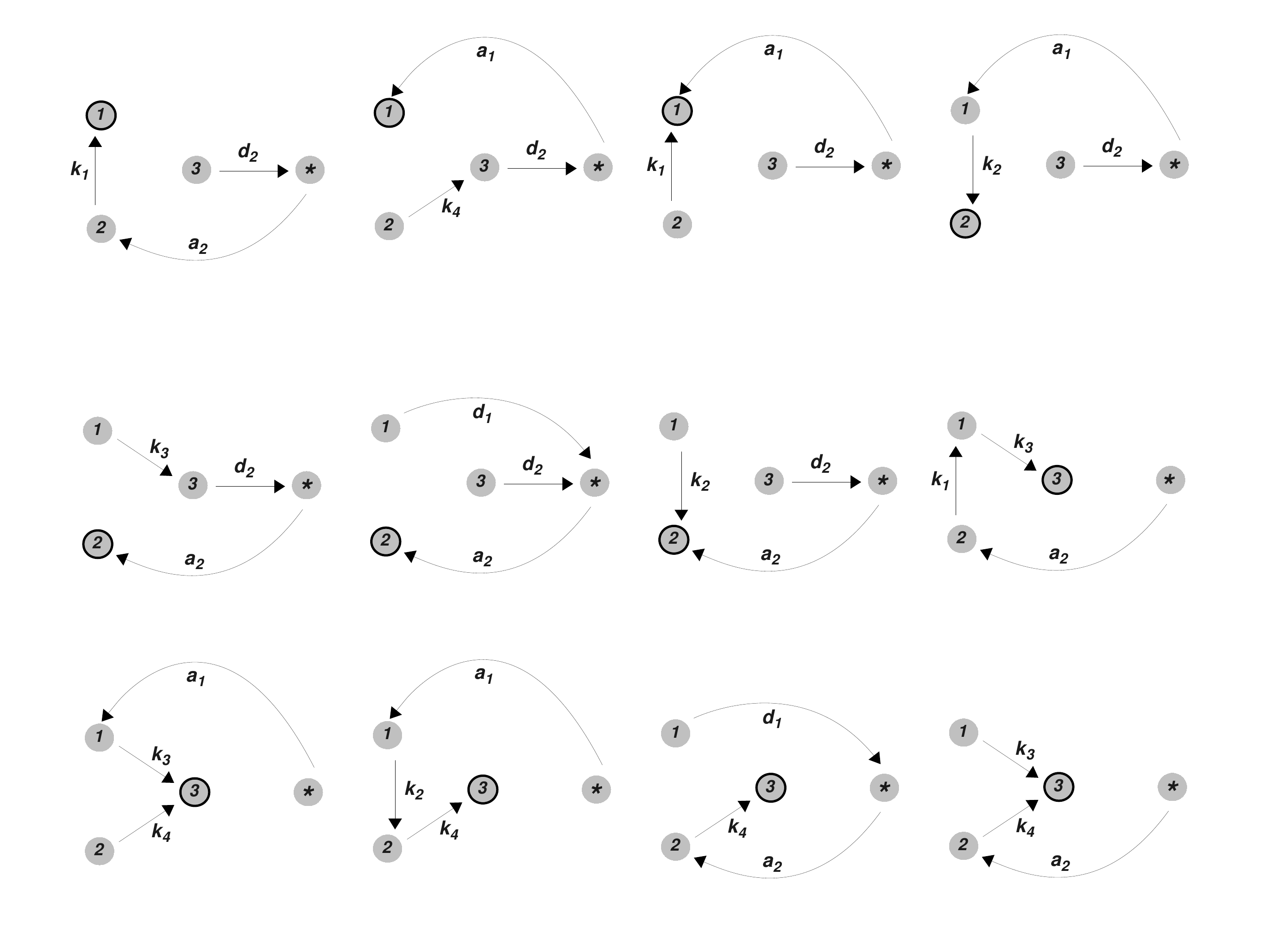}} \\[2.5in]
\subfloat{\includegraphics[viewport=32 230 756 236,width=0.8\textwidth,height=\textheight,keepaspectratio]{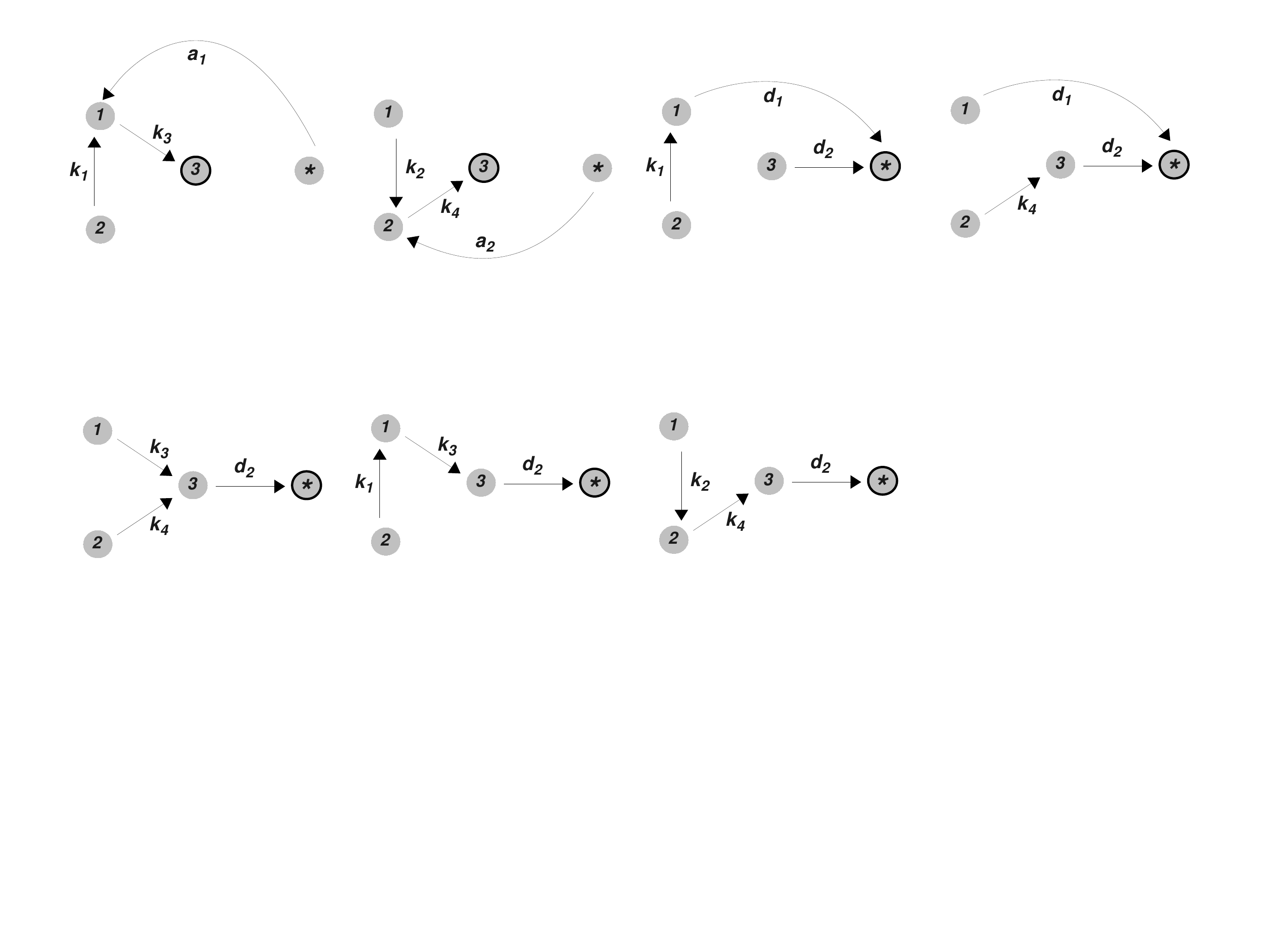}}
\caption{Spanning trees for the labelled, directed graph in Figure~\ref{f-6}B. The 19 trees are listed, with each root indicated by a black circle around the corresponding vertex. The formulas for $\rho_1, \rho_2, \rho_3, \rho_*$ in Figure~\ref{f-6}B can be read off according to the MTT formula in Figure~\ref{f-1}B. \label{fs-2}}
\end{figure}

\end{document}